\documentclass[%
 reprint,superscriptaddress,
 amsmath,amssymb,
 aps,
prl,
]{revtex4-1}
\usepackage{amssymb}
\usepackage{amsmath}
\usepackage{bm}
\usepackage{hyperref}
\usepackage{graphicx}
\usepackage{subfigure}
\usepackage{xcolor}

\def\<{\langle}  %% overiding the original command \<
\def\>{\rangle}  %% overiding the original command \>

\begin{document}
% \preprint{APS/123-QED}
% Use the \preprint command to place your local institutional report
% number in the upper righthand corner of the title page in preprint mode.
% Multiple \preprint commands are allowed.
% Use the 'preprintnumbers' class option to override journal defaults
% to display numbers if necessary
%\preprint{}

%Title of paper
\title{Information geometry under hierarchical quantum measurements}
%\title{The uncertainty relation and Quantum Cram\'er-Rao bound}
\author{Hongzhen Chen}
%\email{hzchen@mae.cuhk.edu.hk}
\affiliation{Department of Mechanical and Automation Engineering, The Chinese University of Hong Kong, Shatin, Hong Kong SAR, P.R. China}

\author{Yu Chen}
%\email{anschen@mae.cuhk.edu.hk}
\affiliation{Department of Mechanical and Automation Engineering, The Chinese University of Hong Kong, Shatin, Hong Kong SAR, P.R. China}
%\author{Zhibo Hou}
%\affiliation{Key Laboratory of Quantum Information,University of Science and Technology of China, CAS, Hefei 230026, P. R. China}
%\affiliation{CAS Center For Excellence in Quantum Information and Quantum Physics}
%\author{Guo-Yong Xiang}
%\email{gyxiang@ustc.edu.cn}
%\affiliation{Key Laboratory of Quantum Information,University of Science and Technology of China, CAS, Hefei 230026, P. R. China}
%\affiliation{CAS Center For Excellence in Quantum Information and Quantum Physics}
%\author{Chuan-Feng Li}
%\affiliation{Key Laboratory of Quantum Information,University of Science and Technology of China, CAS, Hefei 230026, P. R. China}
%\affiliation{CAS Center For Excellence in Quantum Information and Quantum Physics}
%\author{Guang-Can Guo}
%\affiliation{Key Laboratory of Quantum Information,University of Science and Technology of China, CAS, Hefei 230026, P. R. China}
%\affiliation{CAS Center For Excellence in Quantum Information and Quantum Physics}
\author{Haidong Yuan}
\email{hdyuan@mae.cuhk.edu.hk}
\affiliation{Department of Mechanical and Automation Engineering, The Chinese University of Hong Kong, Shatin, Hong Kong SAR, P.R. China}
% repeat the \author .. \affiliation  etc. as needed
% \email, \thanks, \homepage, \altaffiliation all apply to the current
% auth+or. Explanatory text should go in the []'s, actual e-mail
% address or url should go in the {}'s for \email and \homepage.
% Please use the appropriate macro foreach each type of information

% \affiliation command applies to all authors since the last
% \affiliation command. The \affiliation command should follow the
% other information
% \affiliation can be followed by \email, \homepage, \thanks as well.
%\author{Hongzhen Chen}
%\email{hzchen@mae.cuhk.edu.hk}
%\author{Haidong Yuan}
%\homepage[]{Your web page}
%\thanks{}
%\altaffiliation{}
%\affiliation{Department of Mechanical and Automation Engineering,\\The Chinese University of Hong Kong, Shatin, Hong Kong}

%Collaboration name if desired (requires use of superscriptaddress
%option in \documentclass). \noaffiliation is required (may also be
%used with the \author command).
%\collaboration can be followed by \email, \homepage, \thanks as well.
%\collaboration{}
%\noaffiliation

\date{\today}

\begin{abstract}
In most quantum technologies, measurements need to be performed on the parametrized quantum states to transform the quantum information to classical information. The measurements, however, inevitably distort the information. The characterization of the discrepancy is an important subject in quantum information science, which plays a key role in understanding the difference between the structures of the quantum and classical information. Here we analyze the discrepancy in terms of the Fisher information metric and present a framework that can provide analytical bounds on the difference under hierarchical quantum measurements. Specifically, we present a set of analytical bounds on the difference between the quantum and classical Fisher information metric under hierarchical p-local quantum measurements, which are measurements that can be performed collectively on at most p copies of quantum states. The results can be directly transformed to the precision limit in multi-parameter quantum metrology, which leads to characterizations of the tradeoff among the precision of different parameters. The framework also provides a coherent picture for various existing results by including them as special cases. %Our study significantly deepens the understanding of the difference between the quantum and classical information structure, which has wide implications in quantum information science.  %including them as the special cases  %by including them as special cases of the obtained bounds. %are  as special cases and provides a coherent picture on the understanding of the achievable precision under general p-local measurements.
%The information encoded in the parametrized quantum states needs to be extracted by the measurements, which transform quantum states to classical probability distributions. The measurements, however, inevitably distort the information encoded in the quantum states, which is reflected in the change of the information geometry on quantum states and classical probability distributions. In this Letter, we study the change of the Fisher information metric from quantum states to classical probability distributions under general quantum measurements that can be performed collectively on at most p copies of quantum states. We provide analytical bounds on the gap between the quantum Fisher information matrix(QFIM) and the classical Fisher information matrix(CFIM), which constraints the maximal Fisher information that can be extracted from parametrized quantum states by p-local measurements. A variation of the partial commutative condition is identified as the necessary condition for the zero gap between the QFIM and CFIM. We then establish the connection between the partial commutative condition and the weak commutative condition by explicitly proving that the partial commutative condition reduces to the weak commutative condition when $p\rightarrow \infty$. These bounds can be used to characterize the tradeoff among the precision limit for the estimation of multiple parameters, a topic attracted much attention in quantum metrology.
\end{abstract}

% insert suggested PACS numbers in braces on next line
% insert suggested keywords - APS authors don't need to do this
%\keywords{}

%\maketitle must follow title, authors, abstract, \pacs, and \keywords
\maketitle
Quantum measurement serves as the gateway between quantum information and classical information. In most quantum technologies, the information encoded in the parametrized quantum states needs to be extracted by the measurements. For example, in quantum metrology, the estimation of unknown parameters encoded in the quantum states is achieved through the measurements on the parametrized quantum states; in the variational quantum circuit certain information needs to be extracted via the measurements on the parametrized quantum states to update the circuit. Upon the measurements, however, certain properties of quantum information, such as noncommutativity, are lost. This inevitably induces distortions on the information structure. Understanding such distortion is an important subject in quantum information science which helps distinguish the structure of the quantum and classical information. %One way to understand the distortion is the data processing inequality which states that the amount of information can not increase after any channel while general positive operator valued measurement(POVM) can be regarded as a special set of quantum channels( quantum-classical channel).
It also helps to understand the maximal amount of information that can be extracted from quantum states.

In this Letter, we characterize the structure of the information in terms of the Fisher information metric\cite{Hels76book,Hole82book,BrauCM96,GiovLM04,GiovLM06,Haya05book,Barndorff_Nielsen_2000,Escher2011,Naga07beating,HiggBBW07,XianHBW11,Slus17unconditional,daryanoosh2018experimental,yuan2017quantum,yuan2017M,Liu2022} and study the distortion of the metric under hierarchical p-local quantum measurements, which are the measurements that can be performed collectively on at most p copies of quantum states. We note that the Fisher information metric is the only Riemannian metric in information geometry that is invariant under sufficient statistics\cite{Amari2000book}, and has been employed in a broad range of applications\cite{Meyer2021}, such as quantum metrology\cite{Hole82book,Hels76book}, quantum phase transition\cite{doi:10.1142/S0217979210056335,Wang_2014}, entanglement witness\cite{PhysRevA.85.022321,PhysRevA.85.022322}, as well as the natural gradient and effective dimension in statistical learning\cite{JMLR:v21:17-678,Amari2000book}.

We first introduce the quantum and classical Fisher information metric together with the existing results, then present an approach to characterize the achievable classical Fisher information metric under hierarchical p-local quantum measurements. This framework provides a systematical way to generate various analytical bounds on the difference between the quantum and classical Fisher information metric under general p-local measurements. The framework also includes various existing results, which were seemingly disconnected from each other previously, as special cases thus put them into a coherent picture.  %framework with analytical bounds on the difference between the quantum and classical Fisher information metric and show how they include the existing results as special cases.

Given d-dimensional parametrized quantum state, $\rho_x$, with $n$ parameters as $x=(x_1,\cdots, x_n)$, the $jk$-th entry of the quantum Fisher information matrix(QFIM), denoted as $F_Q(x)$, is given by\cite{Hels76book,Hole82book,BrauCM96}
%\begin{equation}
 $   F_Q(x)_{jk}=\frac{1}{2}Tr[\rho_x(L_jL_k+L_kL_j)]$,
%\end{equation}
here $L_j(L_k)$ is the symmetric logarithmic derivative(SLD) with respect to the parameter $x_j(x_k)$ and can be obtained from the equation
%\begin{equation}
 $   \frac{\partial \rho_x}{\partial x_j}=\frac{1}{2}(\rho_x L_j+L_j\rho_x).$
%\end{equation}
The QFIM quantifies the quantum Fisher information metric, which is related to the Bures metric as\cite{BrauCM96}
%\begin{equation}
 $   D_{B}^2(\rho_x,\rho_{x+dx})=\frac{1}{4}dxF_Q(x)dx^T,$
%\end{equation}
here $dx=(dx_1,\cdots, dx_n)$ are infinitesimal changes of the parameters and $D_{B}(\rho_1,\rho_2)=\sqrt{2-2Tr\sqrt{\sqrt{\rho_1}\rho_2\sqrt{\rho_1}}}$ is the Bures distance\cite{Bure69}. The QFIM is additive with respect to the copies of the state, i.e., the QFIM of $\rho_x^{\otimes p}$, which we denote as $F_{Qp}$, equals to $pF_Q$.  %It can be similarly related to the Kullback–Leibler divergence as $D_{KL}(\rho_x\|rho_{x+dx})=\frac{1}{2}dx F_Qdx^T$.
\begin{figure}
  \includegraphics[width=0.3\textwidth]{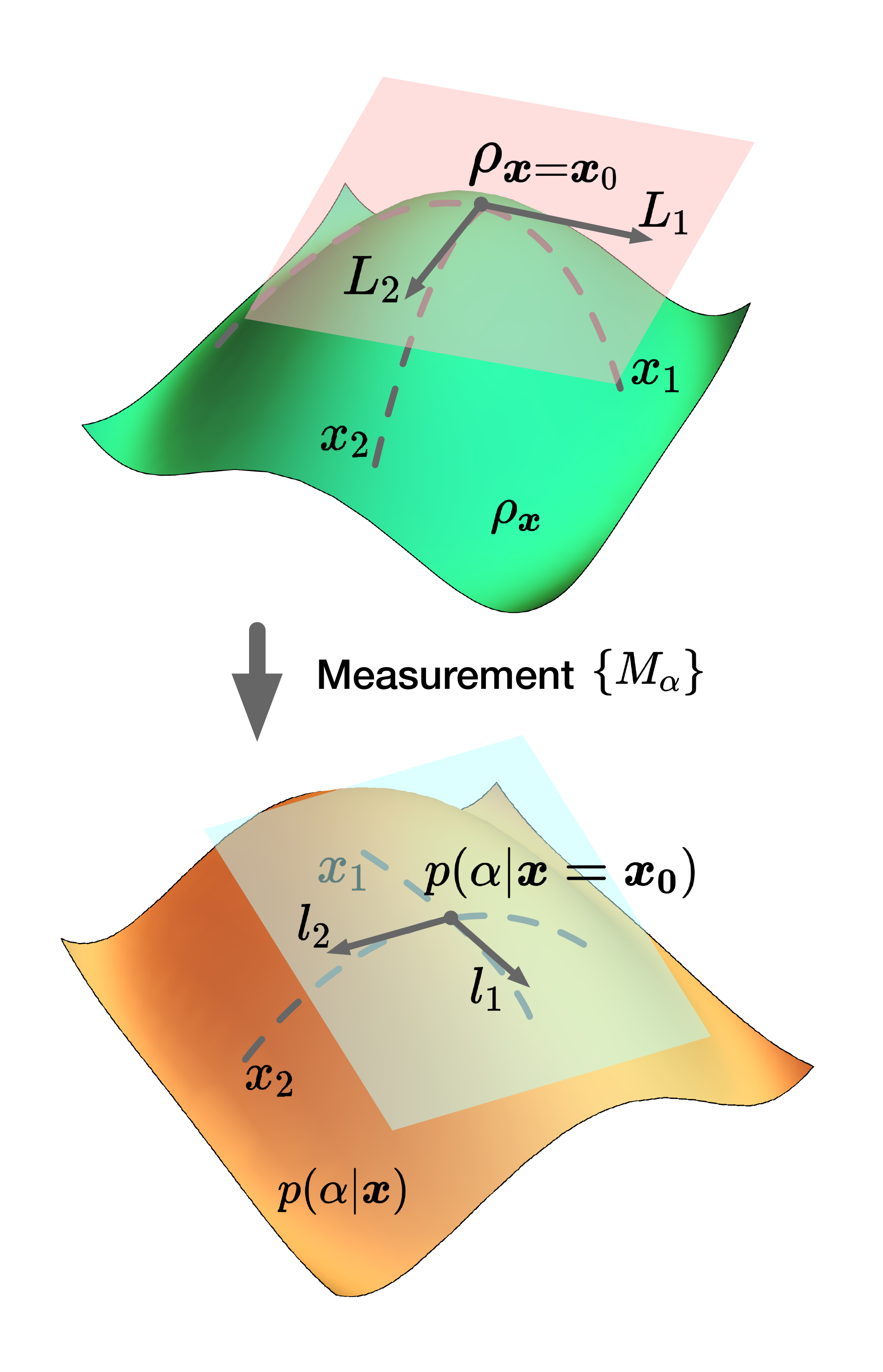}
  \caption{The measurement transforms the parametrized quantum states to classical probability distribution where the classical Fisher information metric is always upper bounded by the quantum Fisher information metric.}
  \label{fig:1copy}
\end{figure}

General measurements that transform the parametrized quantum states to classical data is the positive operator-valued measurement(POVM), which is a set of $\{M_{\alpha}\}$ with $M_\alpha\geq 0$ and $\sum_{\alpha}M_{\alpha}=I$. The probability of obtaining the result $\alpha$ is given by $p(\alpha|x)=Tr(\rho_xM_{\alpha})$. As illustrated in Fig.\ref{fig:1copy}, the measurement changes the parametrized quantum state to the parametrized probability distribution. 

The Fisher information metric of the probability distribution is given by the classical Fisher information matrix(CFIM)\cite{Fish22}, denoted as $F_C(x)$, whose $jk$-th entry is given by
%\begin{equation}
 $   F_C(x)_{jk}=\int_{\alpha} \frac{1}{p(\alpha|x)}\frac{\partial p(\alpha|x) }{\partial x_j}\frac{\partial p(\alpha|x) }{\partial x_k}d\alpha.$
%\end{equation}
The classical Fisher information metric is related to the Euclidean distance between $\sqrt{p(\alpha|x)}$ and $\sqrt{p(\alpha|x+dx)}$ as
\begin{equation}
    \int_{\alpha}(\sqrt{p(\alpha|x)}-\sqrt{p(\alpha|x+dx)})^2d\alpha=\frac{1}{4}dx F_C(x)dx^T.
\end{equation}
We note that for diagonal quantum states, where the diagonal entries can be regarded as the classical probability distribution, the Bures distance reduces to the Euclidean distance.

The QFIM and CFIM characterize the geometrical structure of the parametrized quantum states and the classical probability distribution respectively. As measurements can be regarded as a special set of quantum channels(known as the quantum-classical channel), %$M(\rho_x)=\sum_{\alpha}Tr(M_\alpha\rho_x)|\alpha\rangle\langle\alpha|$ with $\{|\alpha\rangle\}$ as a set of orthogonal states),
the data process inequality tells that the distance can not increase after the measurements. This implies that under any measurement $F_C(x)\leq F_Q(x)$.  %$$dx F_C(x)dx^T\leq dx F_Q(x)dx^T$ for any $dx$, which leads to $F_C(x)\leq F_Q(x)$ under any measurement\cite{Hels76book}.

In classical estimation, the CFIM provides an asymptotically achievable lower bound on the covariance for locally unbiased estimators of the parameters, which is known as the Cramer-Rao bound\cite{Cram46} with
%\begin{equation}
 $   Cov(\hat{x})\geq \frac{1}{\nu}F_C^{-1}(x),$
%\end{equation}
here $\hat{x}$ denotes the locally unbiased estimator of x and $Cov(\hat{x})$ denotes the covariance matrix where the $jk$-th entry is given by $Cov(\hat{x})_{jk}=E[(\hat{x}_j-x_j)(\hat{x}_k-x_k)]$, $\nu$ is the number of sampled data. Since $F_C(x)\leq F_Q(x)$, the covariance matrix is further bounded by the QFIM as
%\begin{equation}
 $   Cov(\hat{x})\geq \frac{1}{\nu}F_Q^{-1}(x),$
%\end{equation}
which is known as the quantum Cramer-Rao bound(QCRB)\cite{Hels76book,Hole82book}. %which, however, is in general not saturable.

\begin{figure}
  \includegraphics[width=0.5\textwidth]{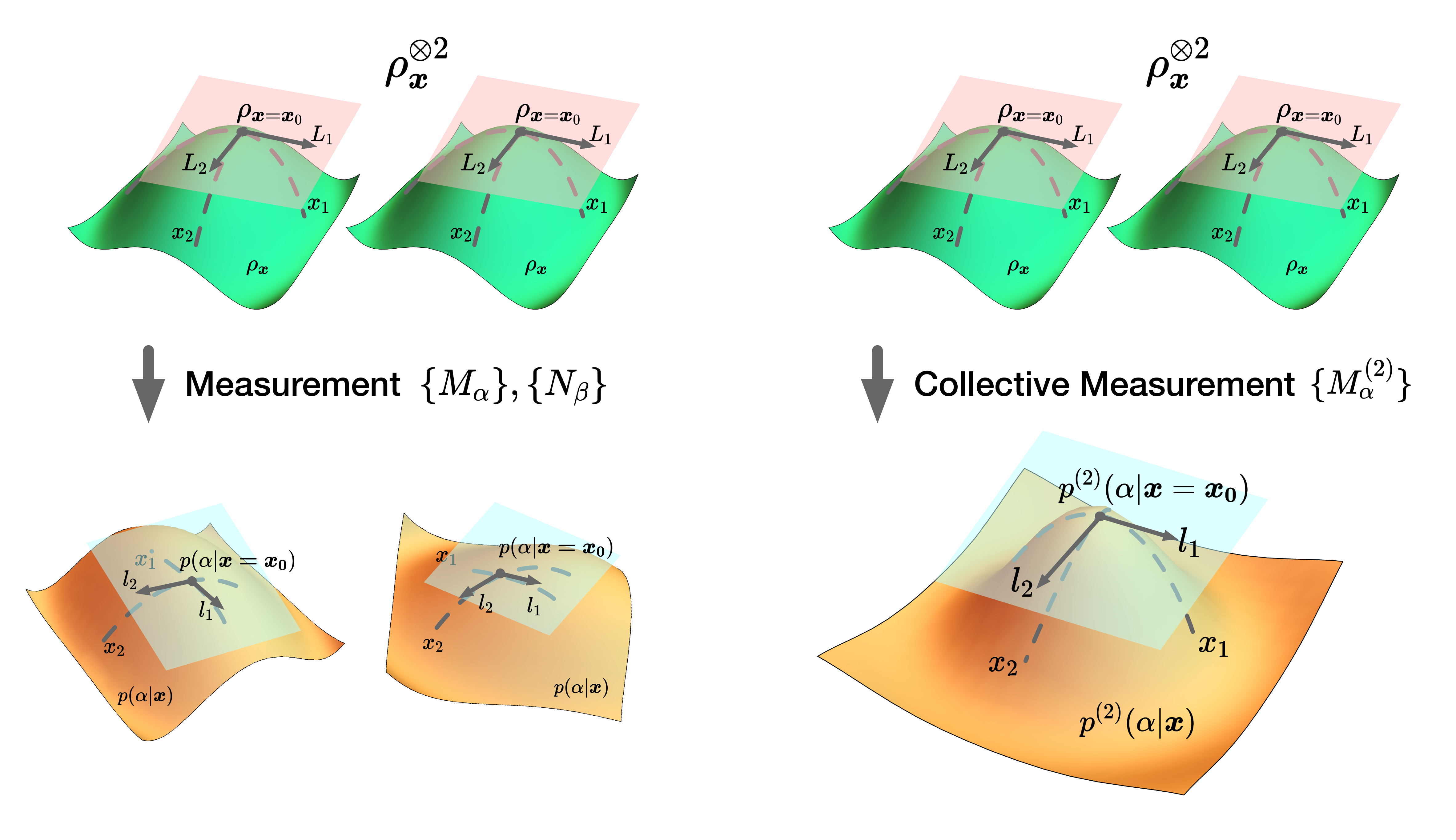}
  \caption{The transformation of information geometry under 1-local and 2-local measurements respectively. Since there are more freedom in the 2-local measurements, the classical Fisher information metric under the optimal 2-local measurement is in general less distorted than the metric under the 1-local measurements, i.e., the classical Fisher information matrix under 2-local measurements can be closer to the quantum Fisher information matrix. The classical Fisher information matrix under $p$-local measurements can get even closer when $p$ increases.}  \label{fig:2copy}
\end{figure}

If there exists a measurement such that $F_C(x)=F_Q(x)$, the measurement then preserves the local geometrical structure and the QCRB is saturable. It is known that if the quantum state is parametrized by a single parameter, i.e., $n=1$, then there always exists a measurement that can preserve the local Fisher information structure \cite{Hels76book}. Furthermore, the measurement that saturates the QCRB can be taken as 1-local measurement, collective measurements are not required. One such measurement is the projective measurement on the eigenvectors of the SLD\cite{Hels76book}. %This is 1-local measurement as the measurement can be performed on one copy of the quantum state, collective measurements on multiple copies of quantum states are not required.
When there are multiple parameters, the information structure becomes much more complicated. First, the SLDs for different parameters may not commute with each other, thus in general there does not exist a measurement that can make $F_C(x)=F_Q(x)$ \cite{GillM00,ALBARELLI2020126311,Carollo_2019,Lu2021,Ragy2016,Chen_2017,Liu_2019,ChenHZ2019,Francesco2019,Rafal2020,Kok2020,vidrighin2014,crowley2014,Yue2014,Zhang2014,Liu2017,Roccia_2017,e22111197,Candeloro2021,Koichi2013,Kahn2009,Yuxiang2019,Suzuki2016,Sidhu2021,Zhu2018universally,Matsumoto_2002,PhysRevLett.111.070403,PhysRevLett.119.130504,Haya05book,Hou20minimal,Federico2021,Houeabd2986,HouSuper2021,Yang2019,PhysRevLett.116.180402,Zhu2018universally,Magdalena2016,PhysRevLett.123.200503}. The distortion of the Fisher information structure is then typically inevitable. Second, the collective measurements matter. As illustrated in Fig.\ref{fig:2copy}, if we repeat the 1-local measurement on two copies of quantum states, the maximal CFIM at most doubles. However, if collective measurements can be performed, it is possible to obtain larger CFIM due to more degrees of freedom in the measurements. In general under p-local measurements, the CFIM that can be extracted from $p$ copies of quantum states, which we denote as $F_{Cp}(x)$, can be larger than $pF_C(x)$. The CFIM is thus super-additive with respect to the number of copies of quantum states. This is related to the notion of the nonlocality without entanglement\cite{Bennett1999}.

%The problem now is to quantify the gap between the QFIM and the CFIM under general p-local measurement.
%In general the amount of the classical Fisher information increases with $p$ when the collectively measurements can be performed on $p$ copies of quantum states.
We use $Tr[F_{Qp}^{-1}(x)F_{Cp}(x)]$ to quantify the difference between the QFIM and the CFIM that can be extracted by $p$-local measurements on $\rho_x^{\otimes p}$\cite{GillM00,ALBARELLI2020126311,Federico2021,Carollo_2019}. Compared to other quantifiers of the difference, such as $\|F_{Qp}(x)-F_{Cp}(x)\|$, $Tr[F_{Qp}^{-1}(x)F_{Cp}(x)]$ has the advantage that it is invariant under reparametrization. Since $F_{Cp}(x)\leq F_{Qp}(x)$, we have a trivial upper bound $Tr[F_{Qp}^{-1}(x)F_{Cp}(x)]\leq Tr(I_n)=n$ ( here $I_n$ is the $n\times n$ Identity matrix), and this upper bound is saturated when there exists a p-local measurement that makes $F_{Cp}(x)=F_{Qp}(x)$. In general $Tr[F_{Qp}^{-1}(x)F_{Cp}(x)]<n$ and  $n-Tr[F_{Qp}^{-1}(x)F_{Cp}(x)]$ quantifies the gap between the QFIM and the achievable CFIM under p-local measurements. %under p-local measurement. %When there is no gap, i.e., , , which equals to $Tr(I_n)=n$ .   %If the parameters are changed to $\tilde{x}$, we have $F_{Qp}(x)=G^TF_{Qp}(\tilde{x})G$ and $F_{Cp}(x)=G^TF_{Cp}(\tilde{x})G$, here $G_{jk}=\frac{\partial \tilde{x}_j}{\partial x_k}$, then $Tr[F_{Qp}^{-1}(x)F_{Cp}(x)]=Tr[F_{Qp}^{-1}(\tilde{x})F_{Cp}(\tilde{x})]$. If
We use $\Gamma_p$ to denote the maximal $Tr[F_{Qp}^{-1}(x)F_{Cp}(x)]$ over all $p$-local measurements, then $\Gamma_1\leq \Gamma_2\leq \cdots \leq \Gamma_{\infty}\leq n$.

Previous results on achievable CFIM are mostly on the extreme cases with $p=1$, $2$ or $p=\infty$\cite{Hole82book,GillM00,Carollo_2019,Lu2021,Ragy2016,Zhu2018universally}. Some of the previous results are stated in terms of the covariance matrix instead of the CFIM, we note that since the classical Cram\'er-Rao bound is achievable asymptotically, the covariance matrix and the inverse of the CFIM is interchangeable. %Here we present a framework for general p-local measurement which bridges the extreme cases. The framework not only connects the existing extreme cases but also provides a versatile tool that leads to various new bounds for general p-local measurement. We now first list the existing results on the extreme p then present the general framework with new bounds and show how it connects and improves the results on the extreme cases. %With this framework, we obtain analytical bounds for general p and show that it not only bridges the existing

For 1-local measurement(p=1), Nagaoka provided a bound for the estimation of two parameters($n=2$) as\cite{Nagaoka1,Nagaoka2}
\begin{equation}
    \aligned
    &\nu Tr[Cov(\hat{x})]\\
    &\geq \min_{\{X_1,X_2\}}Tr(\rho_xX_1^2)+Tr(\rho_xX_2^2)+\|\sqrt{\rho_x}[X_1,X_2]\sqrt{\rho_x}\|_1,
    \endaligned
\end{equation}
here $\{X_j\}$ are Hermitian operators that satisfy the locally unbiased condition, $Tr(\rho_xX_j)=0$ and $Tr(\frac{\partial \rho_x}{\partial x_k}X_j)=\delta_{jk}$ with $\delta_{jk}$ as the Kronecker delta function. Recently Nagaoka bound has been generalized to $n$ parameters \cite{Conlon2021}, which in general can only be evaluated numerically. Gill and Massar provided an analytical bound under 1-local measurements as
    $Tr[F_Q^{-1}(x)F_C(x)]\leq d-1$ with $d$ as the dimension of the Hilbert space for a single $\rho_x$\cite{GillM00}. The Gill-Massar bound is nontrivial only when $n\geq d$.

A necessary condition for the saturation of the QCRB under 1-local measurements is the partial commutative condition\cite{Yang2019}, which requires the SLDs commute on the support of $\rho_x$. Specifically if we write $\rho_x$ in the eigenvalue decomposition as $\rho_x=\sum_{i=1}^m\lambda_i|\Psi_i\rangle\langle\Psi_i|$ with $\lambda_i>0$, the partial commutative condition is $\langle \Psi_r|[L_j,L_k]|\Psi_s\rangle=0$ for any $j,k\in\{1,\cdots, n\}$ and $r,s\in\{1,\cdots, m\}$\cite{Yang2019}.

For p=2, Zhu and Hayashi provided an upper bound on $\Gamma_2$ as $Tr[F_{Q2}^{-1}(x)F_{C2}(x)]\leq \frac{3}{2}(d-1)$\cite{Zhu2018universally}, which is nontrivial only when $n\geq \frac{3}{2}(d-1)$.

For $p=\infty$, Holevo provided an achievable bound\cite{Hole82book} in terms of the weighted covariance matrix as
$\nu Tr[WCov(\hat{x})]\geq \min_{\{X_j\}}\{Tr[W ReZ(X)]+\|\sqrt{W}Im Z(X)\sqrt{W}\|_1\}$, where $W\geq0$ is a weighted matrix, $Z(X)$ is a matrix with its $jk$-th entry given by $Z(X)_{jk}=Tr(\rho_xX_jX_k)$. Here $\{X_1,\cdots, X_n\}$ is a set of Hermitian operators that satisfy the local unbiased condition, %$Tr(\rho_x X_j)=0$ for any $j\in\{1,\cdots, n\}$ and $Tr(\partial_{x_k}\rho_xX_j)=\delta_k^j$ with $\delta_k^j$ as the Kronecker delta, $\delta_k^j=1$ when $k= j$ and $\delta_k^j=0$ when $k\neq j$,
$ReZ(x)$ and $ImZ(X)$ are the real and imaginary part of $Z(x)$ respectively. The Holevo bound in general can only be evaluated numerically \cite{PhysRevLett.123.200503}. For pure states the Holevo bound can be saturated by $1$-local measurements\cite{Matsumoto_2002}, while for mixed states the saturation of the Holevo bound in general requires collective measurements on infinite number of copies of the states. The necessary and sufficient condition for the Holevo bound to coincide with the QCRB is the weak commutative condition, which is $Tr(\rho_x[L_j,L_k])=0$ for all $j, k\in\{1,\cdots, n\}$. %i.e., when the weak commutative condition holds, there exists a measurement under which $\Gamma_\infty=n$. %and the measurement can be 1-local for pure states but in general needs to be performed collectively on infinite copies of quantum states%, that saturates the QCRB and  in this case.
%When the weak commutative condition does not hold, the measurements inevitably distort the Fisher information geometry.

%As the Holevo bound corresponds to the minimal value upon all choice of $\{X_j\}$, by choosing a particular choice of $\{X_j\}$ as $X_j=\sum_k(F_Q^{-1})_{jk}L_k$ and $W=F_Q$, it leads to\cite{Carollo_2019}
%\begin{equation}
%    \nu Tr[F_QCov(\hat{x})]\leq n+\|F_Q^{-\frac{1}{2}}F_{Im}F_Q^{-\frac{1}{2}}\|_1\leq 2n,
%\end{equation}
%here $F_{Im}$ is a matrix with the $jk$-th entry given by $(F_{Im})_{jk}=\frac{1}{2i}Tr(\rho_x[L_j,L_k])$. The last inequality is obtained from the fact that $F_Q+iF_{Im}\geq 0$, which leads to $\|F_Q^{-\frac{1}{2}}F_{Im}F_Q^{-\frac{1}{2}}\|_1\leq \|F_Q^{-\frac{1}{2}}F_QF_Q^{-\frac{1}{2}}\|_1=n$. This provides a lower bound on $\Gamma_{\infty}$ through the Cauchy-Schwarz inequality\cite{Federico2021},
%\begin{eqnarray}\label{eq:CS}
%\aligned
%Tr[F_Q^{\frac{1}{2}}Cov^{\frac{1}{2}}(\hat{x})Cov^{\frac{1}{2}}(\hat{x})F_Q^{\frac{1}{2}}]Tr[F_Q^{-\frac{1}{2}}Cov^{-\frac{1}{2}}(\hat{x})Cov^{-\frac{1}{2}}(\hat{x})F_Q^{-\frac{1}{2}}]
%&Tr[F_QCov(\hat{x})]Tr[F_Q^{-1}Cov^{-1}(\hat{x})]\\
%&\geq |Tr[F_Q^{\frac{1}{2}}Cov^{\frac{1}{2}}(\hat{x})Cov^{-\frac{1}{2}}(\hat{x})F_Q^{-\frac{1}{2}}]|^2=n^2,
%\endaligned
%\end{eqnarray}
%as
%\begin{equation}\label{eq:lowerbound}
%   \Gamma_{\infty}=\frac{1}{\nu} Tr[F_Q^{-1}Cov^{-1}(\hat{x})]\geq \frac{n^2}{n+\|F_Q^{-\frac{1}{2}}F_{Im}F_Q^{-\frac{1}{2}}\|_1}.%\geq \frac{n}{2}.
%\end{equation}
For general p-local measurements, there is little understanding on the achievable CFIM. We present an approach that can lead to various bounds on $\Gamma_p$ for general $p$. These bounds provide a necessary condition for the saturation of the QCRB under general p-local measurements, which recovers the partial commutative condition at p=1 and the weak commutative condition at $p\rightarrow \infty$.
%It also includes the Holevo bound and the Nagaoka bound as special cases, which thus provides a coherent picture for existing seemingly disconnected results.

For a state, $\rho_x$, with $x=(x_1,\cdots, x_n)$, given any POVM, $\{M_\alpha\}$, and any $|u\rangle$, we define $Cov_u$ as a $n\times n$ matrix with the $jk$-th entry given by
\begin{equation}
    (Cov_u)_{jk}=\sum_{\alpha}(\hat{x}_j(\alpha)-x_j)(\hat{x}_k(\alpha)-x_k)\langle u|\sqrt{\rho_x}M_{\alpha} \sqrt{\rho_x}|u\rangle,
\end{equation}
and $A_u$ as a $n\times n$ matrix with the $jk$-th entry given by
\begin{eqnarray}\label{eq:Au}
\aligned
&(A_u)_{jk}=\langle u|\sqrt{\rho_x}X_j X_k\sqrt{\rho_x}|u\rangle\\
&=\frac{1}{2}\langle u|\sqrt{\rho_x}\{X_j,X_k\}\sqrt{\rho_x}|u\rangle+i\frac{1}{2i}\langle u|\sqrt{\rho_x}[X_j,X_k]\sqrt{\rho_x}|u\rangle,
\endaligned
\end{eqnarray}
here $\{\hat{x}_j\}$ are locally unbiased estimators and $\{X_j=\sum_{\alpha}[\hat{x}_j(\alpha)-x_j]M_{\alpha}\}$ satisfy the locally unbiased condition. For any $|u\rangle$ we can prove that $Cov_u\geq A_u$ and $Cov_u\geq A_u^T$. And for any set of $\{|u_q\rangle\}$ that satisfies $\sum_q |u_q\rangle\langle u_q|=I$, it is easy to verify that $Cov(\hat{x})=\sum_q Cov_{u_q}$.  %in particular this holds when $\{|u_q\rangle\}$ form a complete basis. %, $\{|u_1\rangle, \cdots, |u_d\rangle\}$,
%This can be verified by comparing $\sum_q (Cov_{u_q})_{jk}$ and $Cov(\hat{x})_{jk}$ as
%\begin{eqnarray}\label{eq:Covu}
%\aligned
%&\sum_q (Cov_{u_q})_{jk}\\
%&=\sum_q\sum_{\alpha}(\hat{x}_j(\alpha)-x_j)(\hat{x}_k(\alpha)-x_k)\langle u_q|\sqrt{\rho_x}M_{\alpha} \sqrt{\rho_x}|u_q\rangle\\
%&=\sum_{\alpha}(\hat{x}_j(\alpha)-x_j)(\hat{x}_k(\alpha)-x_k)Tr(\rho_xM_{\alpha})\\
%&=Cov(\hat{x})_{jk}.
%\endaligned
%\end{eqnarray}
Then for any choices of $\mathbf{\bar{A}}_{u_q}\in \{A_{u_q}, A^T_{u_q}\}$, we have %$Cov_{u_q}\geq \mathbf{\bar{A}}_{u_q}$ and
%\begin{equation}
$Cov(\hat{x})=\sum_q Cov_{u_q} \geq \mathbf{\bar{A}}=\sum_q\mathbf{\bar{A}}_{u_q},$
%\end{equation}
here each $\mathbf{\bar{A}}_{u_q}$ can be taken independently as either $A_{u_q}$ or $A^T_{u_q}$.
By decompose $\mathbf{\bar{A}}$ into the real and imaginary part as  $\bold{\bar{A}}=\bold{\bar{A}}_{Re}+i\bold{\bar{A}}_{Im}$, we obtain a bound on the weighted covariance matrix,
\begin{equation}\label{eq:bound}
   \nu Tr[WCov(\hat{x})]\geq \min_{\{X_j\}}Tr[W\bold{\bar{A}}_{Re}]+\|\sqrt{W}\bold{\bar{A}}_{Im}\sqrt{W}\|_1,
    \end{equation}
    where the number of repetition, $\nu$, is included. Any choices of $\{|u_q\rangle\}$ with $\sum_q |u_q\rangle\langle u_q|=I$ and any $\mathbf{\bar{A}}_{u_q}\in \{A_{u_q}, A^T_{u_q}\}$ lead to a valid bound.

This provides a versatile tool to obtain many useful bounds by properly choosing $\{|u_q\rangle\}$ and $\{\bar{\bold{A}}_{u_q}\}$. In particular the Holevo bound\cite{Hole82book} and the Nagaoka bound\cite{Nagaoka1,Nagaoka2} can be recovered from this general bound by making particular choices of $\{|u_q\rangle\}$ and $\{\bar{\bold{A}}_{u_q}\}$\cite{CCYpra}. Furthermore, by combining with an improved Robertson's uncertainty relation for multiple observables\cite{CCYpra,PhysRev.46.794,Trifonov:2002aa}, we can obtain a set of analytical bounds on the gap between the QFIM and the CFIM under general p-local measurements. Specifically for pure states we have
%In this case we have
%\begin{eqnarray}
%\aligned
%\bold{\bar{A}}&=\sum_q \bold{\bar{A}}_{u_q}\\
%&=\begin{pmatrix}
%      Tr(\rho_xX_1^2) & \frac{1}{2}Tr[\rho_x\{X_1,X_2\}]\\
%      \frac{1}{2}Tr[\rho_x\{X_1,X_2\}] & Tr(\rho_xX_2^2)
%    \end{pmatrix}\\
%    &+i\begin{pmatrix}
%      0 & \frac{1}{2}\|\sqrt{\rho_x}[X_1,X_2]\sqrt{\rho_x}\|_1\\
%      -\frac{1}{2}\|\sqrt{\rho_x}[X_1,X_2]\sqrt{\rho_x}\|_1 & 0
%    \end{pmatrix}.
%\endaligned
%\end{eqnarray}
%Eq.(\ref{eq:bound}) then becomes (with $W=I$)
%\begin{eqnarray}
%\aligned
%     &\nu Tr[Cov(\hat{x})]\\
%     &\geq \min_{\{X_1,X_2\}}Tr[\bold{\bar{A}}_{Re}]+\|\bold{\bar{A}}_{Im}\|_1\\
%          &=\min_{\{X_1,X_2\}}Tr(\rho_xX_1^2)+Tr(\rho_xX_2^2)+\|\sqrt{\rho_x}[X_1,X_2]\sqrt{\rho_x}\|_1,
%     \endaligned
%\end{eqnarray}
%which recovers the Nagaoka bound\cite{Nagaoka1,Nagaoka2}.
%The optimal choice of $|u_q\rangle$ and $\mathbf{\bar{A}}_{u_q}$ lead to the tightest bound, but any other choices also lead to valid bounds. %although the optimal choice can be hard to identify,
%By making particular choices, we can generate various analytical upper bounds on $\Gamma_p$. %as listed below.%
  %\footnote{the detailed derivation is in the longer version of the paper submitted together to PRA}.
%\begin{enumerate}
%    \item
    \begin{eqnarray}
    \aligned
    \Gamma\leq &n-f(n)\|F_{Q}^{-\frac{1}{2}}F_{Im}F_{Q}^{-\frac{1}{2}}\|_F^2,
    %&\frac{1}{\nu}Tr[F_Q^{-1}Cov^{-1}(\hat{x})]\\
%    \Gamma\leq &n-\frac{1}{4(n-1)}\|F_{Q}^{-\frac{1}{2}}F_{Im}F_{Q}^{-\frac{1}{2}}\|_F^2,\\
%    \Gamma\leq & n-\frac{n-2}{(n-1)^2}\|F_{Q}^{-\frac{1}{2}}F_{Im}F_{Q}^{-\frac{1}{2}}\|_F^2,\\
 %   \Gamma\leq & n-\frac{1}{5}\|F_{Q}^{-\frac{1}{2}}F_{Im}F_{Q}^{-\frac{1}{2}}\|_F^2,
    \endaligned
    \end{eqnarray}
    here $F_{Im}$ is the matrix with the entries given by $(F_{Im})_{jk}=\frac{1}{2i}Tr(\rho_x[L_j,L_k])$ and %$n$ is the number of parameters
    $\|\cdot{}\|_F=\sqrt{\sum_{j,k}|(\cdot)_{jk}|^2}$ is the Frobenius norm, $f(n)$ can take $\frac{1}{4(n-1)}$, $\frac{n-2}{(n-1)^2}$ or $\frac{1}{5}$, which all lead to valid bounds. Since larger f(n) leads to tighter bound, we can take $f(n)=\max\{\frac{1}{4(n-1)},\frac{n-2}{(n-1)^2},  \frac{1}{5}\}$. Note that here we use $\Gamma$ instead of $\Gamma_p$ since for pure states $\Gamma_1=\Gamma_2=\cdots=\Gamma_{\infty}$. 

 %   \item For pure states, when $n\geq 3$ the bound can be tightened as
  %  \begin{eqnarray}
  %  \aligned
  %          &\frac{1}{\nu}Tr[F_Q^{-1}Cov^{-1}(\hat{x})]\\%&\leq n-\frac{n-2}{(n-1)^2}\|\tilde{F}_{Im}\|_F^2\\
%\leq & n-\frac{n-2}{(n-1)^2}\|F_{Q}^{-\frac{1}{2}}F_{Im}F_{Q}^{-\frac{1}{2}}\|_F^2.
%\endaligned
%    \end{eqnarray}
%And we also have
%    \begin{eqnarray}
%    \aligned
%    \frac{1}{\nu}Tr[F_Q^{-1}Cov^{-1}(\hat{x})]
%    \leq n-\frac{1}{5}\|F_{Q}^{-\frac{1}{2}}F_{Im}F_{Q}^{-\frac{1}{2}}\|_F^2,
%\endaligned
%    \end{eqnarray}
%    which is tighter than the above bound for $n\geq 5$.

%   \item For mixed states under the 1-local measurement, we have
%   \begin{equation}
%    \frac{1}{\nu}Tr[F_Q^{-1}Cov^{-1}(\hat{x})]\leq n-\frac{1}{4(n-1)} \|C_{1}\|_F^2,
%\end{equation}
%with the entries of $C_1$ given by
%\begin{eqnarray}
%\aligned
%$(C_1)_{jk}%&=\frac{1}{2}\|\sqrt{\rho_x}[\tilde{L}_j,\tilde{L}_k]\sqrt{\rho_x}\|_1\\
%=\frac{1}{2}\|\sqrt{\rho_x}[\sum_q (F_Q^{-\frac{1}{2}})_{jq}L_q,\sum_q (F_Q^{-\frac{1}{2}})_{kq}L_q]\sqrt{\rho_x}\|_1.$
%\endaligned
%\end{eqnarray}
%\item
For mixed states under p-local measurements we have %with the $p$-local measurement, when $n\geq 3$ we have
\begin{eqnarray}\label{eq:Fimp}
\aligned
\Gamma_p  &\leq n-f(n)\|\frac{F_Q^{-\frac{1}{2}}\mathbf{\bar{F}}_{Imp}F_Q^{-\frac{1}{2}}}{p}\|_F^2,
  %&\frac{1}{\nu}Tr[F_Q^{-1}Cov^{-1}(\hat{x})]\\
%   \Gamma_p  &\leq n-\frac{n-2}{(n-1)^2}\|\frac{F_Q^{-\frac{1}{2}}\mathbf{\bar{F}}_{Imp}F_Q^{-\frac{1}{2}}}{p}\|_F^2,\\
%\Gamma_p    &\leq n-\frac{1}{5}\|\frac{F_Q^{-\frac{1}{2}}\mathbf{\bar{F}}_{Imp}F_Q^{-\frac{1}{2}}}{p}\|_F^2,
   %Tr(\tilde{C}_1^T\tilde{C}_1),
    \endaligned
\end{eqnarray}
%and
%\begin{eqnarray}
%\aligned
%  \frac{1}{\nu}Tr[F_Q^{-1}Cov^{-1}(\hat{x})]
%     \leq n-\frac{1}{5}\|\frac{F_Q^{-\frac{1}{2}}\mathbf{\bar{F}}_{Imp}F_Q^{-\frac{1}{2}}}{p}\|_F^2, %Tr(\tilde{C}_1^T\tilde{C}_1),
%    \endaligned
%\end{eqnarray}
where $f(n)=\max\{\frac{1}{4(n-1)},\frac{n-2}{(n-1)^2},  \frac{1}{5}\}$, $\mathbf{\bar{F}}_{Imp}$ is the imaginary part of  $\mathbf{\bar{F}}=\sum_q \bar{F}_{u_q}$ with each $\bar{F}_{u_q}$ equal to either $F_{u_q}$ or $F_{u_q}^T$, here $F_{u_q}$ is a $n\times n$ matrix with the $jk$-th entry given by
%\begin{equation}
 $   (F_{u_q})_{jk}=\langle u_q|\sqrt{\rho_x^{\otimes p}}L_{jp}L_{kp}\sqrt{\rho_x^{\otimes p}}|u_q\rangle,$
%\end{equation}
 $L_{jp}$ is the SLD of $\rho_x^{\otimes p}$ corresponding to the parameter $x_j$, $\{|u_q\rangle\}$ are any set of vectors in $H_d^{\otimes p}$ that satisfies $\sum_q |u_q\rangle\langle u_q|=I_{d^p}$ with $I_{d^p}$ denoting the $d^p\times d^p$ Identity matrix. %By making choices of $\{|u_q\rangle\}$ and $\{\bar{F}_{u_q}\}$ various analytical bounds can be obtained.

We can also obtain additional bounds by combining different choices of $\{|u_q\rangle\}$. In particular for mixed states under $p$-local measurements we can get
\begin{eqnarray}\label{eq:boundCp}
\aligned
   % \frac{1}{\nu}Tr[F_Q^{-1}Cov^{-1}(\hat{x})]
    \Gamma_p\leq n-\frac{1}{4(n-1)} \|\frac{C_{p}}{p}\|_F^2, %Tr(\tilde{C}_1^T\tilde{C}_1),
    \endaligned
\end{eqnarray}
here $C_p$ is a $n\times n$ matrix with the $jk$-th entry given by
%\begin{equation}%\label{eq:CP}
  $  (C_p)_{jk}=\frac{1}{2}\|\sqrt{\rho_x^{\otimes p}}[\tilde{L}_{jp}, \tilde{L}_{kp}]\sqrt{\rho_x^{\otimes p}}\|_1,$ 
%\end{equation}
here $\tilde{L}_{jp}(\tilde{L}_{kp})$ is the SLD for $\rho_{\tilde{x}}^{\otimes p}$ with respect to the parameter $\tilde{x}_j(\tilde{x}_k)$, where $\tilde{x}=F_Q^{\frac{1}{2}}x$ is a reparametrization under which $\tilde{F}_Q=I$. Note that we always have $\|\frac{C_{p}}{p}\|_F\geq \|\frac{F_Q^{-\frac{1}{2}}\mathbf{\bar{F}}_{Imp}F_Q^{-\frac{1}{2}}}{p}\|_F$, the bound in Eq.(\ref{boundCp}) is thus tighter than the bound in Eq.(\ref{eq:Fimp}) when $f(n)=\frac{1}{4(n-1)}$, but it could be less tight when $f(n)=\frac{n-2}{(n-1)^2}$ or $\frac{1}{5}$. %$\tilde{L}_{jp}=\sum_{r=1}^p \tilde{L}_j^{(r)}$, here $\tilde{L}_j^{(r)}=I^{\otimes (r-1)}\otimes \tilde{L}_j\otimes I^{\otimes (p-r)}$, $r=1,\cdots, p$, which can also be equivalently written as
%$\tilde{L}_{jp}$ is the SLD of $\rho_x^{\otimes p}$ under the reparametrization such that the QFIM of $\rho_x$ equals to the Identity, specifically
%$\tilde{L}_{jp}=\sum_q (F_Q^{-\frac{1}{2}})_{jq}L_{qp}$ where $(F_Q^{-{\frac{1}{2}}})_{jq}$ is the jq-th entry of $F_Q^{-{\frac{1}{2}}}$ and $L_{qp}$ is the SLD for $\rho_x^{\otimes p}$ with respect to the parameter $x_q$. %$\frac{(C_p)_{jk}}{p}$ decreases monotonically with $p$ as $(C_1)_{jk}\geq \frac{(C_2)_{jk}}{2} \geq \frac{(C_3)_{jk}}{3}\geq \cdots$.

The bound in Eq.(\ref{eq:boundCp}) provides a necessary condition for the saturation of the QCRB under $p$-local measurements, which is $\frac{C_p}{p}=0$. For $p=1$, this condition is equivalent to the partial commutative condition. For $p\rightarrow \infty$, we prove that\cite{CCYpra}
%\begin{equation}
    %(C_1)_{jk}\geq \frac{(C_2)_{jk}}{2}\geq\cdots
% $   \frac{(C_p)_{jk}}{p}\geq \frac{(C_{p+1})_{jk}}{p+1}$ %\geq \cdots
%\end{equation}
%and
\begin{equation}
\lim_{p\rightarrow \infty}\frac{(C_p)_{jk}}{p}=\frac{1}{2}|Tr(\rho_x[\tilde{L}_j,\tilde{L}_k])|.
\end{equation}
%The partial commutative condition, $\frac{C_p}{p}=0$,
At $p\rightarrow \infty$ the condition is thus equivalent to the weak commutative condition, $Tr(\rho_x[\tilde{L}_j,\tilde{L}_k])=0$, $\forall j, k$. This builds a bridge between the partial commutative condition and the weak commutative condition at the two extremes.

% when $p\rightarrow \infty$.
%Here $\tilde{L}_j$ is the SLD under the reparameterization which makes the QFIM equals to the Identity, i.e., $\tilde{L}_j=\sum_q (F_Q^{-\frac{1}{2}})_{jq}L_{q}$ with $L_{q}$ as the SLD of $\rho_x$ corresponding to the original parameter $x_q$. This clarifies the relation between the partial commutative condition and the weak commutative condition.
%The weak commutative condition under this parametrization is equivalent to the weak commutative condition in the original parametrization, $Tr(\rho_x[L_j,L_k])=0$, $\forall j, k$. %Here the weak commutative condition is expressed under the parametrization that $\tilde{F}_Q=I$, which is equivalent to the weak commutative condition in the original parametrization, $Tr(\rho_x[L_j,L_k])=0$, $\forall j, k$ .

%\item

%\item %A set of alternative tradeoff relations are obtained.
The bound in Eq.(\ref{eq:boundCp}) involves operators on $p$ copies of quantum states, whose dimension grows exponentially with $p$. We provide another simpler bound which only uses operators on a single $\rho_x$ as
\begin{equation}
   % \frac{1}{\nu}Tr[F_Q^{-1}Cov^{-1}(\hat{x})]
    \Gamma_p\leq n-\frac{1}{4(n-1)} \|\frac{T_p}{p}\|_F^2,
\end{equation}
where $T_p$ can be computed from the eigenvalues, eigenvectors and  SLDs of a single $\rho_x$. Specifically, given $\rho_x=\sum_{i=1}^m\lambda_i |\Psi_i\rangle\langle\Psi_i|$ in the eigenvalue decomposition with $\lambda_i>0$, the $jk$-th entry of $T_p$ is given by
%$T_p$ is a $n\times n$ matrix with the $jk$-th entry given by
\begin{eqnarray}
\aligned
   (T_p)_{jk}=\frac{1}{2}E(|\sum_{r=1}^p \langle\Phi_{r}|[\tilde{L}_j,\tilde{L}_k]|\Phi_{r}\rangle|),     %&(T_p)_{jk}%=\frac{1}{2}\|D_p^{(jk)}\|_1\\
    %=&\frac{1}{2}\sum_{v_1,\cdots, v_p%\in\{1,\cdots, m\}
    %}(\prod_{r=1}^p\lambda_{v_r}) |\sum_{r=1}^p \langle\Psi_{v_r}|[\tilde{L}_j,\tilde{L}_k]|\Psi_{v_r}\rangle|,
    \endaligned
\end{eqnarray}
here $E(\cdot)$ denotes the expected value, each $|\Phi_r\rangle$ is randomly and independently chosen from the eigenvectors of $\rho_x$ with the probability equal to the corresponding eigenvalue, i.e., each $|\Phi_r\rangle$ takes $|\Psi_i\rangle$ with probability $\lambda_i$, $i\in \{1,\cdots, m\}$. $\tilde{L}_{j}(\tilde{L}_{k})$ is the SLD for $\rho_{\tilde{x}}$ with respect to the parameter $\tilde{x}_j(\tilde{x}_k)$, where $\tilde{x}=F_Q^{\frac{1}{2}}x$.   
%$\tilde{L}_{j}=\sum_{q}(F_Q^{-{\frac{1}{2}}})_{jq}L_{q}$ and $\tilde{L}_{k}=\sum_{q}(F_Q^{-{\frac{1}{2}}})_{kq}L_{q}$, where $L_q$ is the SLD of $\rho_x$ with respect to the parameter $x_q$. %This only uses the eigenvalue decomposition and the SLDs of a single $\rho_x$. % $(T_p)_{jk}$ basically equals to the expected value of $\frac{1}{2}|\sum_{r=1}^p \langle\Psi_{v_r}|[\tilde{L}_j,\tilde{L}_k]|\Psi_{v_r}\rangle|$ with each eigenvector $|\Psi_{v_r}\rangle$ selected with probability $\lambda_{v_r}$.
%$T_p$ can be computed with operators on a single $\rho_x$.
%Although this bound only uses operators on the Hilbert space of a single $\rho_x$,
The difference between this bound and the bound in Eq.(\ref{eq:boundCp} is at most $O(\frac{1}{\sqrt{p}})$. Thus when the bound in Eq.(\ref{eq:boundCp} is hard to compute at large p, we can use this bound instead which is almost as tight.  %Specifically we have

The upper bounds on $\Gamma_p$ can be directly transformed to the lower bounds on the covariance matrix. For example, suppose a p-local measurement is repeated with $\mu$ times(so total $\nu=\mu p$ copies of $\rho_x$), from the classical Cram\'er-Rao bound we have $Cov(\hat{x})\geq \frac{1}{\mu}F_{Cp}^{-1}(x)$ (here the equality is achievable since the classical Cram\'er-Rao bound is saturable). This implies that $\frac{1}{\mu}Cov^{-1}(\hat{x})\leq F_{Cp}(x)$. Any upper bound, $\Gamma_p\leq D$, then leads to an upper bound on $\frac{1}{\nu}Tr[F_Q^{-1}Cov^{-1}(\hat{x})]$ as  $\frac{1}{\nu}Tr[F_Q^{-1}Cov^{-1}(\hat{x})]\leq Tr(F_{Qp}^{-1}F_{Cp})\leq D$(note $F_{Qp}=pF_Q$). %This can be further used to derived analytical lower bounds on $Tr[WCov(\hat{x})]$ via
From the Cathy-Schwartz inequality,
%\begin{equation}
 $   Tr[WCov(\hat{x})]Tr[F_Q^{-1}Cov^{-1}(\hat{x})]\geq (Tr\sqrt{F_Q^{-\frac{1}{2}}WF_Q^{-\frac{1}{2}}})^2,$
%\end{equation}
%the upper bounds $\frac{1}{\nu}Tr[F_Q^{-1}Cov^{-1}(\hat{x})]\leq C$ then leads to
%we can obtain lower bounds under p-local measurements
we then obtain
\begin{equation}
    \nu Tr[WCov(\hat{x})]\geq \frac{(Tr\sqrt{F_Q^{-\frac{1}{2}}WF_Q^{-\frac{1}{2}}})^2}{D}.
\end{equation}
By substituting $D$ with any of the upper bounds on $Tr(F_{Qp}^{-1}F_{Cp})$ obtained above we then get analytical bounds on the weighted covariance matrix.

\emph{Summary}  We provided a framework to quantify the difference between the quantum and classical Fisher information metric under hierarchical quantum measurements. The framework provides a systematic way to generate bounds on the achievable CFIM for general quantum states under general measurements, which significantly improves our understandings on the Fisher information geometry under hierarchical quantum measurements. A necessary condition for the zero gap between the quantum and classical Fisher metric has also been identified, which is shown to recover the partial commutative condition at $p=1$ and the weak commutative condition at $p\rightarrow \infty$. The result can be directly transformed to the precision limits in multi-parameter quantum metrology and have implications in various other fields\cite{Meyer2021, Liu_2019,Haug2021}. The detailed derivation can be found in the companion paper\cite{CCYpra} which also contains additional bounds that can be obtained with the framework. %While we focused on the Fisher information geometry here, the developed approach can also be used to improve the Robertson's uncertainty relations for multiple noncommutative observables and provide bounds on the incompatibility of general noncommutative observables, which deepens our understandings on a distinguish feature of quantum mechanics.   %The obtained results are expected to have wide implication for the study of the precision limit in multi-parameter quantum metrology, the boundary between the quantum and classical information structure, noncommutatively and uncertainty relations. %can be directly used to characterize the precision for the estimation of multiple parameters, a central topic in quantum metrology.

%The parametrized quantum states need to be transformed to classical data via measurements. Such measurements, however, inevitably distort the information encoded in the quantum states. The characterization of the measurement distortion is an important subject in quantum information science, as it helps to distinguish the difference between quantum and classical information. We provided a new framework to characterize the difference in terms of the Fisher information metric. The framework can not only include the existing Holevo bound and the Nagaoka bound, but also lead to various useful new bounds. We demonstrated the power of the framework by deriving hierarchical analytical bounds on the difference between the quantum and classical Fisher information matrix under general p-local measurements which bridges the existing extreme cases.
%For finite $p$, the necessary and sufficient condition for the zero gap remains unknown.  Holevo

%When $p\rightarrow \infty$, i.e., collectively measurements on infinite number of quantum states can be performed, $Tr[F_{Qp}^{-1}(x)F_{Cp}(x)]$

\begin{acknowledgements}
This work is partially supported by the Research Grants Council of Hong Kong with the Grant No. 14307420.
\end{acknowledgements}
%\bibliographystyle{apsrev4-1}
%\bibliographystyle{unsrt}
%\bibliography{reference}

\begin{thebibliography}{10}

\bibitem{Hels76book}
Carl~W. Helstrom.
\newblock {\em Quantum Detection and Estimation Theory}.
\newblock Academic Press, New York, 1976.

\bibitem{Hole82book}
A.~S. Holevo.
\newblock {\em Probabilistic and Statistical Aspects of Quantum Theory}.
\newblock North-Holland, Amsterdam, 1982.

\bibitem{BrauCM96}
Samuel~L. Braunstein, Carlton~M. Caves, and G.~J. Milburn.
\newblock Generalized uncertainty relations: Theory, examples, and {Lorentz}
  invariance.
\newblock {\em Ann. Phys.}, 247(1):135--173, 1996.

\bibitem{GiovLM04}
Vittorio Giovannetti, Seth Lloyd, and Lorenzo Maccone.
\newblock Quantum-enhanced measurements: Beating the standard quantum limit.
\newblock {\em Science}, 306:1330, 2004.

\bibitem{GiovLM06}
Vittorio Giovannetti, Seth Lloyd, and Lorenzo Maccone.
\newblock Quantum metrology.
\newblock {\em Phys. Rev. Lett.}, 96:010401, 2006.

\bibitem{Haya05book}
M.~Hayashi, editor.
\newblock {\em Asymptotic Theory of Quantum Statistical Inference}.
\newblock World Scientific, Singapore, 2005.

\bibitem{Barndorff_Nielsen_2000}
O~E Barndorff-Nielsen and R~D Gill.
\newblock Fisher information in quantum statistics.
\newblock {\em Journal of Physics A: Mathematical and General},
  33(24):4481--4490, jun 2000.

\bibitem{Escher2011}
B~M Escher, R~L {de Matos Filho}, and L~Davidovich.
\newblock {General framework for estimating the ultimate precision limit in
  noisy quantum-enhanced metrology}.
\newblock {\em Nat. Phys.}, 7(5):406--411, may 2011.

\bibitem{Naga07beating}
Tomohisa Nagata, Ryo Okamoto, Jeremy~L O'brien, Keiji Sasaki, and Shigeki
  Takeuchi.
\newblock Beating the standard quantum limit with four-entangled photons.
\newblock {\em Science}, 316(5825):726--729, 2007.

\bibitem{HiggBBW07}
B.~L. Higgins, D.~W. Berry, S.~D. Bartlett, H.~M. Wiseman, and G.~J. Pryde.
\newblock Entanglement-free {Heisenberg}-limited phase estimation.
\newblock {\em Nature}, 450(7168):393--396, 2007.

\bibitem{XianHBW11}
G.~Y. Xiang, B.~L. Higgins, D.~W. Berry, H.~M. Wiseman, and G.~J. Pryde.
\newblock Entanglement-enhanced measurement of a completely unknown optical
  phase.
\newblock {\em Nat. Photonics}, 5:43, 2011.

\bibitem{Slus17unconditional}
Sergei Slussarenko, Morgan~M Weston, Helen~M Chrzanowski, Lynden~K Shalm,
  Varun~B Verma, Sae~Woo Nam, and Geoff~J Pryde.
\newblock Unconditional violation of the shot-noise limit in photonic quantum
  metrology.
\newblock {\em Nature Photonics}, 11(11):700, 2017.

\bibitem{daryanoosh2018experimental}
Shakib Daryanoosh, Sergei Slussarenko, Dominic~W Berry, Howard~M Wiseman, and
  Geoff~J Pryde.
\newblock Experimental optical phase measurement approaching the exact
  heisenberg limit.
\newblock {\em Nature communications}, 9(1):4606, 2018.

\bibitem{yuan2017quantum}
Haidong Yuan and Chi-Hang~Fred Fung.
\newblock Quantum parameter estimation with general dynamics.
\newblock {\em NPJ Quantum Information}, 3(1):14, 2017.

\bibitem{yuan2017M}
Haidong Yuan and Chi-Hang~Fred Fung.
\newblock Quantum metrology matrix.
\newblock {\em Phys. Rev. A}, 96 (1), 012310, 2017.

\bibitem{Liu2022}
Jing Liu, Mao Zhang, Hongzhen Chen, Lingna Wang, and Haidong Yuan.
\newblock Optimal scheme for quantum metrology.
\newblock {\em Adv. Quantum Technol}, 5:2100080, 2022.

\bibitem{Amari2000book}
Shun-ichi Amari and Horishi Nagaoka.
\newblock {\em Methods of Information Geometry}.
\newblock Oxford University Press, New York, 2000.

\bibitem{Meyer2021}
Johannes~Jakob Meyer.
\newblock Fisher {I}nformation in {N}oisy {I}ntermediate-{S}cale {Q}uantum
  {A}pplications.
\newblock {\em {Quantum}}, 5:539, September 2021.

\bibitem{doi:10.1142/S0217979210056335}
SHI-JIAN GU.
\newblock Fidelity approach to quantum phase transitions.
\newblock {\em International Journal of Modern Physics B}, 24(23):4371--4458,
  2010.

\bibitem{Wang_2014}
Teng-Long Wang, Ling-Na Wu, Wen Yang, Guang-Ri Jin, Neill Lambert, and Franco
  Nori.
\newblock Quantum fisher information as a signature of the superradiant quantum
  phase transition.
\newblock {\em New Journal of Physics}, 16(6):063039, jun 2014.

\bibitem{PhysRevA.85.022321}
Philipp Hyllus, Wies\l{}aw Laskowski, Roland Krischek, Christian Schwemmer,
  Witlef Wieczorek, Harald Weinfurter, Luca Pezz\'e, and Augusto Smerzi.
\newblock Fisher information and multiparticle entanglement.
\newblock {\em Phys. Rev. A}, 85:022321, Feb 2012.

\bibitem{PhysRevA.85.022322}
G\'eza T\'oth.
\newblock Multipartite entanglement and high-precision metrology.
\newblock {\em Phys. Rev. A}, 85:022322, Feb 2012.

\bibitem{JMLR:v21:17-678}
James Martens.
\newblock New insights and perspectives on the natural gradient method.
\newblock {\em Journal of Machine Learning Research}, 21(146):1--76, 2020.

\bibitem{Bure69}
D~J~C Bures.
\newblock An extension of {Kakutani's} theorem on infinite product measures to
  tensor product of semifinite $w^*$-algebras.
\newblock {\em Trans. Am. Math. Soc.}, 135:199--212, 1969.

\bibitem{Fish22}
R.~A. Fisher.
\newblock On the mathematical foundations of theoretical statistics.
\newblock {\em Philos. Trans. R. Soc. Lond. A}, 222:309--368, 1922.

\bibitem{Cram46}
Harald Cram\'er.
\newblock {\em Mathematical Methods of Statistics}.
\newblock Princeton University Press, Princeton, NJ, 1946.

\bibitem{GillM00}
Richard~D. Gill and Serge Massar.
\newblock State estimation for large ensembles.
\newblock {\em Phys. Rev. A}, 61:042312, Mar 2000.

\bibitem{ALBARELLI2020126311}
F.~Albarelli, M.~Barbieri, M.G. Genoni, and I.~Gianani.
\newblock A perspective on multiparameter quantum metrology: From theoretical
  tools to applications in quantum imaging.
\newblock {\em Physics Letters A}, 384(12):126311, 2020.

\bibitem{Carollo_2019}
Angelo Carollo, Bernardo Spagnolo, Alexander~A Dubkov, and Davide Valenti.
\newblock On quantumness in multi-parameter quantum estimation.
\newblock {\em Journal of Statistical Mechanics: Theory and Experiment},
  2019(9):094010, sep 2019.

\bibitem{Lu2021}
Xiao-Ming Lu and Xiaoguang Wang.
\newblock Incorporating heisenberg's uncertainty principle into quantum
  multiparameter estimation.
\newblock {\em Phys. Rev. Lett.}, 126:120503, Mar 2021.

\bibitem{Ragy2016}
Sammy Ragy, Marcin Jarzyna, and Rafal Demkowicz-Dobrza\'{n}ski.
\newblock Compatibility in multiparameter quantum metrology.
\newblock {\em Phys. Rev. A}, 94:052108, Nov 2016.

\bibitem{Chen_2017}
Yu~Chen and Haidong Yuan.
\newblock Maximal quantum fisher information matrix.
\newblock {\em New Journal of Physics}, 19(6):063023, jun 2017.

\bibitem{Liu_2019}
Jing Liu, Haidong Yuan, Xiao-Ming Lu, and Xiaoguang Wang.
\newblock Quantum fisher information matrix and multiparameter estimation.
\newblock {\em Journal of Physics A: Mathematical and Theoretical},
  53(2):023001, dec 2020.

\bibitem{ChenHZ2019}
Hongzhen Chen and Haidong Yuan.
\newblock Optimal joint estimation of multiple rabi frequencies.
\newblock {\em Phys. Rev. A}, 99:032122, Mar 2019.

\bibitem{Francesco2019}
Francesco Albarelli, Marco Barbieri, Marco~G. Genoni, and Ilaria Gianani.
\newblock A perspective on multiparameter quantum metrology: From theoretical
  tools to applications in quantum imaging.
\newblock {\em Phys. Lett. A}, 384:126311, 2020.

\bibitem{Rafal2020}
R.~Demkowicz-Dobrza\'{n}ski, W.~G\'{o}recki, and M.~Gu\c{t}\u{a}.
\newblock Multi-parameter estimation beyond quantum fisher information.
\newblock {\em J. Phys. A: Math. Theor.}, 53:363001, 2020.

\bibitem{Kok2020}
Jasminder~S. Sidhu and Pieter Kok.
\newblock Geometric perspective on quantum parameter estimation.
\newblock {\em AVS Quantum Sci.}, 2:014701, 2020.

\bibitem{vidrighin2014}
M.~D. Vidrighin, G.~Donati, M.~G. Genoni, X.-M. Jin, W.~S. Kolthammer, M.~S.
  Kim, A.~Datta, M.~Barbieri, and I.~A. Walmsley.
\newblock Joint estimation of phase and phase diffusion for quantum metrology.
\newblock {\em Nat. Commun.}, 5:3532, 2014.

\bibitem{crowley2014}
Philip J.~D. Crowley, Animesh Datta, Marco Barbieri, and I.~A. Walmsley.
\newblock Tradeoff in simultaneous quantum-limited phase and loss estimation in
  interferometry.
\newblock {\em Phys. Rev. A}, 89:023845, Feb 2014.

\bibitem{Yue2014}
J.-D. Yue, Y.-R. Zhang, and H~Fan.
\newblock Quantum-enhanced metrology for multiple phase estimation with noise.
\newblock {\em Sci. Rep.}, 4:5933, 2014.

\bibitem{Zhang2014}
Yu-Ran Zhang and Heng Fan.
\newblock Quantum metrological bounds for vector parameters.
\newblock {\em Phys. Rev. A}, 90:043818, Oct 2014.

\bibitem{Liu2017}
Jing Liu and Haidong Yuan.
\newblock Control-enhanced multiparameter quantum estimation.
\newblock {\em Phys. Rev. A}, 96:042114, Oct 2017.

\bibitem{Roccia_2017}
Emanuele Roccia, Ilaria Gianani, Luca Mancino, Marco Sbroscia, Fabrizia Somma,
  Marco~G Genoni, and Marco Barbieri.
\newblock Entangling measurements for multiparameter estimation with two
  qubits.
\newblock {\em Quantum Science and Technology}, 3(1):01LT01, nov 2017.

\bibitem{e22111197}
Sholeh Razavian, Matteo G.~A. Paris, and Marco~G. Genoni.
\newblock On the quantumness of multiparameter estimation problems for qubit
  systems.
\newblock {\em Entropy}, 22(11), 2020.

\bibitem{Candeloro2021}
Alessandro Candeloro, Matteo~G.A. Paris, and Marco~G. Genoni.
\newblock On the properties of the asymptotic incompatibility measure in
  multiparameter quantum estimation.
\newblock {\em arxiv}, page 2107.13426, 2021.

\bibitem{Koichi2013}
Koichi Yamagata, Akio Fujiwara, and Richard~D. Gill.
\newblock {Quantum local asymptotic normality based on a new quantum likelihood
  ratio}.
\newblock {\em The Annals of Statistics}, 41(4):2197 -- 2217, 2013.

\bibitem{Kahn2009}
Jonas Kahn and Mădălin Guţă.
\newblock Local asymptotic normality for finite dimensional quantum systems.
\newblock {\em Communications in Mathematical Physics}, 289:597--652, Jul 2009.

\bibitem{Yuxiang2019}
Yuxiang. Yang, Giulio Chiribella, and Masahito Hayashi.
\newblock Attaining the ultimate precision limit in quantum state estimation.
\newblock {\em Communications in Mathematical Physics}, 368:223--293, 2019.

\bibitem{Suzuki2016}
Jun Suzuki.
\newblock Explicit formula for the holevo bound for two-parameter qubit-state
  estimation problem.
\newblock {\em Journal of Mathematical Physics}, 57(4):042201, 2016.

\bibitem{Sidhu2021}
Jasminder~S. Sidhu, Yingkai Ouyang, Earl~T. Campbell, and Pieter Kok.
\newblock Tight bounds on the simultaneous estimation of incompatible
  parameters.
\newblock {\em Phys. Rev. X}, 11:011028, Feb 2021.

\bibitem{Zhu2018universally}
Huangjun Zhu and Masahito Hayashi.
\newblock Universally fisher-symmetric informationally complete measurements.
\newblock {\em Phys. Rev. Lett.}, 120:030404, Jan 2018.

\bibitem{Matsumoto_2002}
K~Matsumoto.
\newblock A new approach to the cram{\'{e}}r-rao-type bound of the pure-state
  model.
\newblock {\em Journal of Physics A: Mathematical and General},
  35(13):3111--3123, mar 2002.

\bibitem{PhysRevLett.111.070403}
Peter~C. Humphreys, Marco Barbieri, Animesh Datta, and Ian~A. Walmsley.
\newblock Quantum enhanced multiple phase estimation.
\newblock {\em Phys. Rev. Lett.}, 111:070403, Aug 2013.

\bibitem{PhysRevLett.119.130504}
Luca Pezz\`e, Mario~A. Ciampini, Nicol\`o Spagnolo, Peter~C. Humphreys, Animesh
  Datta, Ian~A. Walmsley, Marco Barbieri, Fabio Sciarrino, and Augusto Smerzi.
\newblock Optimal measurements for simultaneous quantum estimation of multiple
  phases.
\newblock {\em Phys. Rev. Lett.}, 119:130504, Sep 2017.

\bibitem{Hou20minimal}
Zhibo Hou, Zhao Zhang, Guo-Yong Xiang, Chuan-Feng Li, Guang-Can Guo, Hongzhen
  Chen, Liqiang Liu, and Haidong Yuan.
\newblock Minimal tradeoff and ultimate precision limit of multiparameter
  quantum magnetometry under the parallel scheme.
\newblock {\em Phys. Rev. Lett.}, 125:020501, Jul 2020.

\bibitem{Federico2021}
Federico Belliardo and Vittorio Giovannetti.
\newblock Incompatibility in quantum parameter estimation.
\newblock {\em New Journal of Physics}, 23(6):063055, jun 2021.

\bibitem{Houeabd2986}
Zhibo Hou, Jun-Feng Tang, Hongzhen Chen, Haidong Yuan, Guo-Yong Xiang,
  Chuan-Feng Li, and Guang-Can Guo.
\newblock Zero{\textendash}trade-off multiparameter quantum estimation via
  simultaneously saturating multiple heisenberg uncertainty relations.
\newblock {\em Science Advances}, 7(1), 2021.

\bibitem{HouSuper2021}
Zhibo Hou, Yan Jin, Hongzhen Chen, Jun-Feng Tang, Chang-Jiang Huang, Haidong
  Yuan, Guo-Yong Xiang, Chuan-Feng Li, and Guang-Can Guo.
\newblock "super-heisenberg" and heisenberg scalings achieved simultaneously in
  the estimation of a rotating field.
\newblock {\em Phys. Rev. Lett.}, 126:070503, Feb 2021.

\bibitem{Yang2019}
Jing Yang, Shengshi Pang, Yiyu Zhou, and Andrew~N. Jordan.
\newblock Optimal measurements for quantum multiparameter estimation with
  general states.
\newblock {\em Phys. Rev. A}, 100:032104, Sep 2019.

\bibitem{PhysRevLett.116.180402}
Nan Li, Christopher Ferrie, Jonathan~A. Gross, Amir Kalev, and Carlton~M.
  Caves.
\newblock Fisher-symmetric informationally complete measurements for pure
  states.
\newblock {\em Phys. Rev. Lett.}, 116:180402, May 2016.

\bibitem{Magdalena2016}
Magdalena Szczykulska, Tillmann Baumgratz, and Animesh Datta.
\newblock Multi-parameter quantum metrology.
\newblock {\em Advances in Physics: X}, 1(4):621--639, 2016.

\bibitem{PhysRevLett.123.200503}
Francesco Albarelli, Jamie~F. Friel, and Animesh Datta.
\newblock Evaluating the holevo cram\'er-rao bound for multiparameter quantum
  metrology.
\newblock {\em Phys. Rev. Lett.}, 123:200503, Nov 2019.

\bibitem{Bennett1999}
Charles~H. Bennett, David~P. DiVincenzo, Christopher~A. Fuchs, Tal Mor, Eric
  Rains, Peter~W. Shor, John~A. Smolin, and William~K. Wootters.
\newblock Quantum nonlocality without entanglement.
\newblock {\em Phys. Rev. A}, 59:1070--1091, Feb 1999.

\bibitem{Nagaoka1}
H.~Nagaoka.
\newblock A new approach to cramer–rao bounds for quantum state estimation.
\newblock In Masahito Hayashi, editor, {\em Asymptotic theory of quantum
  statistical inference: Selected Papers}, Singapore, 2005. World scientific.
\newblock Originally published as IEICE Technical Report, 89, 228, IT 89–42,
  9–14 (1989).

\bibitem{Nagaoka2}
H.~Nagaoka.
\newblock A generalization of the simultaneous diagonalization of hermitian
  matrices and its relation to quantum estimation theory.
\newblock In Masahito Hayashi, editor, {\em Asymptotic theory of quantum
  statistical inference: Selected Papers}, Singapore, 2005. World scientific.

\bibitem{Conlon2021}
Lorcán~O. Conlon, Jun. Suzuki, Ping~Koy Lam, and Syed~M. Assad.
\newblock Efficient computation of the nagaoka–hayashi bound for
  multiparameter estimation with separable measurements.
\newblock {\em npj Quantum Information}, 7:110, Jul 2021.

\bibitem{CCYpra}
Hongzhen Chen, Yu Chen and Haidong Yuan.
\newblock Incompatibility measures in multi-parameter quantum estimation under hierarchical quantum measurements.
\newblock {\em to appear in Physical Review A}, 2022. arXiv:2109.05807.

\bibitem{PhysRev.46.794}
H.~P. Robertson.
\newblock An indeterminacy relation for several observables and its classical
  interpretation.
\newblock {\em Phys. Rev.}, 46:794--801, Nov 1934.

\bibitem{Trifonov:2002aa}
D.~A. Trifonov.
\newblock Generalizations of heisenberg uncertainty relation.
\newblock {\em The European Physical Journal B - Condensed Matter and Complex
  Systems}, 29(2):349--353, 2002.

\bibitem{Haug2021}
Tobias Haug, Kishor Bharti, and M.S. Kim.
\newblock Capacity and quantum geometry of parametrized quantum circuits.
\newblock {\em PRX Quantum}, 2:040309, Oct 2021.

\end{thebibliography}

\end{document}